\documentclass[a4paper]{article}

\usepackage{amsmath, amsfonts, amssymb, theorem, verbatim}

\allowdisplaybreaks[1]

\numberwithin{equation}{section}

\newtheorem{theorem}{Theorem}[section]
\newtheorem{corollary}[theorem]{Corollary}
\newtheorem{lemma}[theorem]{Lemma}
\newtheorem{proposition}[theorem]{Proposition}
{\theorembodyfont{\rmfamily} 
\newtheorem{remark}[theorem]{Remark}} 
{\theorembodyfont{\rmfamily} 
} 
{\theorembodyfont{\rmfamily} 
\newtheorem{definition}[theorem]{Definition}}

{\theorembodyfont{\rmfamily} 
}

\makeatletter

\newenvironment{proof}{\noindent\textsc{Proof.}}
{\hspace*{\fill}$\square$\par\bigskip}

\def\section{\@startsection {section}{1}{\z@}{3.5ex plus 1ex minus
    .2ex}{2.3ex plus .2ex}{\large\bf}}
    \def\subsection{\@startsection{subsection}{2}{\z@}{3.25ex plus 1ex minus 
 .2ex}{1.5ex plus .2ex}{\bf}}

\makeatother

\title{On the symmetry of commuting differential operators with
singularities along hyperplanes} 
\author{Kenji Taniguchi 
\thanks{Department of Mathematics, Aoyama Gakuin University, 
5-10-1, Fuchinobe, Sagamihara, Kanagawa 229-8558, Japan. 
(taniken@gem.aoyama.ac.jp) }}

\date{\empty}

\begin{document}

\maketitle

\begin{abstract}
We study the commutants of a Schr\"{o}dinger operator 
whose potential function possesses inverse square singularities along
some hyperplanes passing through the origin. 
It is shown that the Weyl group symmetry of the potential function and
the commutants naturally results from such singularities and
the generic nature of the coupling constants. 

\end{abstract}


\section{Introduction}

The Calogero-Moser-Sutherland models and their
generalizations developed by Olshanetsky and Perelomov are completely
integrable systems with long-range interactions. 
These systems are closely related to root systems and Weyl
groups. 
Let $(\Sigma, W)$ be a pair consisting of a root system and the
corresponding Weyl group. 
In the quantum case, 
the Schr\"{o}dinger operator is
\begin{equation}\label{eq:CMS-Schrodinger} 
- \Delta 
+ \sum_{\alpha \in \Sigma^{+}} 
m_{\alpha} (m_{\alpha} + 1) 
\langle \alpha, \alpha \rangle 
u(\langle \alpha, x \rangle), 
\end{equation}
where $\Delta = \sum_{i=1}^{n} \partial^2/\partial x_{i}^{2}$, 
$\langle u, v \rangle$ is the standard inner
product of $u, v \in \boldsymbol{R}^{n}$, 
the quantities $m_{\alpha}$ are $W$-invariant parameters, 
and 
$u(t) = t^{-2}$ (rational case), 
$\omega^{2} \sinh^{-2} \omega t$, 
$\omega^{2} \sin^{-2} \omega t$ (trigonometric cases) or
$\wp(t)$ (elliptic case). 
Obviously, this operator is $W$-invariant. 
In addition to this operator, there are well-known conserved operators
for such a system that are $W$-invariant. 
In the rational potential case, we can consider integrable
systems invariant under the action of finite Coxeter
groups.

It is evident that the potential function of the above Schr\"{o}dinger
operator possesses inverse-square singularities along the reflection
hyperplanes of $W$. 
The main object of this paper is to show that the Weyl group (or
Coxeter group) symmetry of such a system results naturally from the
inverse square singularities and the generic nature of the parameters
$m_{\alpha}$.

To make the following discussion more precise, we introduce some
notation. 
For a non-zero vector $\alpha \in \boldsymbol{R}^{n}$, 
we denote by $H_{\alpha}$ the hyperplane 
$\langle \alpha, x \rangle = 0$ and by $r_{\alpha}$ reflection with
respect to $H_{\alpha}$. 
For a finite set $\mathcal{H}$ of mutually non-parallel
vectors in $\boldsymbol{R}^{n}$, let $L$ be the Schor\"{o}dinger 
operator defined by 
\begin{align}\label{eq:Schrodinger}
L &= 
- \Delta 
+ R(x), 
&
R(x) 
&= 
\sum_{\alpha \in \mathcal{H}} 
\frac{C_{\alpha}}{\langle \alpha, x \rangle^{2}} 
+ 
\widetilde{R}(x), 
\end{align}
where 
$\widetilde{R}(x)$ is real analytic at $x=0$ 
and the constants $C_{\alpha}$ are non-zero for $\alpha \in
\mathcal{H}$. 
We call $\mathcal{H}$ the {\it hyperplane arrangement} of $L$ or
the {\it hyperplane arrangement} of $R(x)$. 

Assume that there exists a commutant $P$ of $L$ with constant 
principal symbol. 
Note that we do {\it not} assume the symmetry of either $R(x)$ or $P$, 
nor do we assume $\mathcal{H}$ to be a subset of a root system. 
The first result is stated as follows.

\begin{theorem}\label{theorem:reflection invariance}  
If $C_{\alpha} \not= k(k+1) \langle \alpha, \alpha \rangle$ for
any integer $k$, then the principal symbol of $P$ is
$r_{\alpha}$-invariant. 
\end{theorem} 
We prove this theorem in Sections 
\ref{section:Rank one reduction} and 
\ref{section:The rank one rational case}. 

We call the potential function $R(x)$ {\it generic} if
$C_{\alpha} \not= k(k+1) \langle \alpha, \alpha \rangle$ for any
integer $k$ and for any $\alpha \in \mathcal{H}$. 
In non-generic cases, many interesting phenomena have
been observed. 
For example, if the parameters $m_{\alpha}$ are integers, 
then there exist $W$-non-invariant conserved
operators for \eqref{eq:CMS-Schrodinger}, in addition to the 
$W$-invariant ones \cite{CV, VSC}. 
Also in non-generic cases, Veselov, Feigin and Chalykh found new
completely integrable systems like
\eqref{eq:CMS-Schrodinger}, 
but whose hyperplane arrangements are not root systems but deformed
root systems \cite{CFV, VFC}. 

Though non-generic cases like those mentioned above are very
interesting, 
we restrict our attention to generic cases beginning in Section 
\ref{section:Hyperplane arrangement in a generic case}. 
In Section 
\ref{section:Hyperplane arrangement in a generic case}, 
we address the problem of determining the permissible kinds of
hyperplanes arrangements. 
To avoid unnecessary complication, 
we assume the ``irreducibility'' of $\mathcal{H}$
(Definition~\ref{def:irreducibility of H}). 
The main result in Section 
\ref{section:Hyperplane arrangement in a generic case} is
that if $\mathcal{H}$ is irreducible, the potential function
is generic, and $L$ has a non-trivial commutant, 
then $\mathcal{H}$ must be a subset of the positive root
system of some finite reflection group 
(Theorem~\ref{thm:main in section 4}). 

In Sections 
\ref{section:Determination of the potential -- general
setting}, 
\ref{section:Determination of the potential -- type
$A_{n-1}$} 
and \ref{section:Determination of the potential 
-- type $B_{n}$ and $D_{n}$}, 
we determine the potential function $R(x)$ in the case that the
root system containing $\mathcal{H}$ is of the classical type. 
The type $A$ case is treated in Section 
\ref{section:Determination of the potential -- type
$A_{n-1}$}, 
and the types $B$ and $D$ are treated in Section 
\ref{section:Determination of the potential 
-- type $B_{n}$ and $D_{n}$}. 
We now give a brief summary of the arguments given in those sections. 
In the case that the root system $\Sigma$
containing $\mathcal{H}$ is of type $A$, $B$ or $D$, 
we assume that the Schr\"{o}dinger operator
\eqref{eq:Schrodinger} commutes with a differential operator
$P$, whose principal symbol is 
$\sum_{i<j<k} \xi_{i} \xi_{j} \xi_{k}$ for type $A$ and $\sum_{i<j}
\xi_{i}^{2} \xi_{j}^{2}$ for types $B$ and $D$. 
Under this assumption, we can show that $\mathcal{H}$ must coincide
with $\Sigma^{+}$ and that the potential function $R(x)$ must be Weyl
group invariant. 

In \cite{OO}, \cite{OOS} and \cite{OS}, 
Ochiai, Oshima and Sekiguchi classified the potential functions
$R(x)$ satisfying the relation $[-\Delta + R(x), P]=0$, 
which do not necessarily possess poles along hyperplanes, in the Weyl
group invariant context. 
They also proved that $-\Delta + R(x)$ is completely integrable for
such $R(x)$. 
In Sections 6 and 7, it is shown that our potential function $R(x)$
and the commutant $P$ are identical to those that they considered. 
Therefore, it is seen that $R(x)$ is one of the functions classified
in \cite{OOS} (Theorem~\ref{thm:A-type last}, 
Remark~\ref{rem:BD-type last}), 
and $L$ is completely integrable. 
Hence, the complete integrability of $L$ essentially
follows from the generic nature of coupling constants and the
existence of one non-trivial commutant $P$. 

\noindent
\textbf{Acknowledgements.} 
The author would like to thank Professor H. Ochiai for helpful
discussions on some points of this paper. 
He also thanks a referee for carefully considered and kind comments. 
This reserch was partially supported by Grant-in-Aid for Scientific
Research (C)(2) No.\,15540183, Japan Society for the Promotion of
Science.



\section{Rank-one reduction}
\label{section:Rank one reduction}
To begin, we introduce some notation. 
Let $\{e_{1}, \dots, e_{n}\}$ be the standard basis of
$\boldsymbol{R}^{n}$ and $x = (x_{1}, \dots, x_{n})$ be the
corresponding coordinates. 
For simplicity, denote by $\partial_{i}$ the partial differential
$\partial/\partial x_{i}$ and define 
$\partial = (\partial_{1}, \dots, \partial_{n})$. 
We denote the norm of a vector $v \in \boldsymbol{R}^{n}$ by $|v|$. 
An $m_{0}$-th order diferential operator $P$ is expressed as 
\[
P= 
\sum_{k = 0}^{m_{0}} P_{k}, 
\qquad 
P_{k} = 
\sum_{|p| = m_{0} - k} a_{p}(x) \partial^{p}, 
\]
where $p = (p_{1}, \dots, p_{n}) \in \boldsymbol{N}^{n}$ is a
multi-index,  
and $|p|$ is the length $\sum_{i} p_{i}$ of $p$. 
Corresponding to this operator, 
we introduce 
\begin{align*}
\widetilde{P}_{k} 
&= 
\sum_{|p| = m_{0} - k} a_{p}(x) \xi^{p} 
&
&\mbox{and}
&
\widetilde{P} 
&= 
\sum_{k=0}^{m_{0}} \widetilde{P}_{k}
& 
(\xi 
&= (\xi_{1}, \dots, \xi_{n})),  
\end{align*}
and call them the {\it symbols} of $P_{k}$ and $P$, respectively. 
In particular, $\widetilde{P}_{0}$ is called the 
{\it principal symbol} of $P$. 

In this section, we reduce the proof of
Theorem~\ref{theorem:reflection invariance} to that for the rank-one
rational case. 
For this purpose, 
we introduce a new coordinate system on $\boldsymbol{R}^{n}$. 
First, choose $\alpha \in \mathcal{H}$. 
Then, let 
$e_{1}' 
= |\alpha|^{-1} \alpha$ and  let 
$\{e_{2}', \dots, e_{n}'\}$ be an orthonormal basis of
$H_{\alpha}$. 
The corresponding coordinates are denoted  
$y = (y_{1}, \dots, y_{n})$. 
It is easy to see that 
$\langle \alpha, x \rangle 
= |\alpha| y_{1}$ 
and 
$\Delta = \sum_{i=1}^{n} \partial_{y_{i}^{2}}$. 
In terms of these new coordinates, 
$L$ and $P$ are expressed as 
\begin{align}
L &= -\sum_{i=1}^{n} \partial_{y_{i}}^{2} 
+ \frac{\langle \alpha, \alpha \rangle^{-1} C_{\alpha}}{y_{1}^{2}} 
+ S(y), 
\notag \\
P &= 
\sum_{k=0}^{m_{0}} P_{k},
\qquad  
P_{k} = 
\sum_{|p| = m_{0} - k} b_{p}(y) \partial_{y}^{p}, 
\label{eq:P_k}
\end{align} 
where $\partial_{y} = (\partial_{y_{1}}, \dots, \partial_{y_{n}})$, 
and $S(y)$ is a real analytic function on 
$D = \{|y| < \epsilon\} 
\setminus 
\cup_{\beta \in \mathcal{H} \atop \beta \not= \alpha} 
H_{\beta}$ for some $\epsilon > 0$. 
Next, let $\eta = (\eta_{1}, \dots, \eta_{n})$ be the symbol corresponding
to $\partial_y = (\partial_{y_{1}}, \dots, \partial_{y_{n}})$. 
Thus, we denote the symbol of $P_{k}$ given in \eqref{eq:P_k} by 
\[
\widetilde{P}_{k}(y, \eta)  
= 
\sum_{|p| = m_{0} - k} b_{p}(y) \eta^{p}. 
\]
By the Leibniz rule, we have 
\[
(P_{k} \circ y_{1}^{-2}) \widetilde{\enskip}
= 
\sum_{l=0}^{m_{0} - k} (-1)^{l} (l+1) y_{1}^{-l-2} 
\partial_{\eta_{1}}^{l} \widetilde{P}_{k}. 
\]
Therefore, we obtain 
\begin{align*} 
[P, y_{1}^{-2}] \widetilde{\enskip} 
&= \sum_{k=0}^{m_{0}-1} \sum_{l=1}^{m_{0}-k} (-1)^{l} (l+1)
y_{1}^{-l-2} \partial_{\eta_{1}}^{l} \widetilde{P}_{k} 
\\
&= \sum_{k=1}^{m_{0}} \sum_{l=1}^{k} (-1)^{l} (l+1)
y_{1}^{-l-2} \partial_{\eta_{1}}^{l} \widetilde{P}_{k-l}. 
\end{align*}
On the other hand, because 
\begin{equation*} 
[\Delta, P] \widetilde{\enskip} 
= 
2 \sum_{k=0}^{m_{0}} \langle \eta, \partial_{y} \rangle 
\widetilde{P}_{k} 
+ \sum_{k=0}^{m_{0}} \Delta \widetilde{P}_{k}  
\qquad 
(\langle \eta, \partial_{y} \rangle 
= 
\sum_{i = 1}^{n} \eta_{i} \partial_{y_{i}}), 
\end{equation*} 
we have 
\begin{align}
& [L, P] = 0 
\notag\\
& \Leftrightarrow 
2 \sum_{k=0}^{m_{0}} \langle \eta, \partial_{y} \rangle 
\widetilde{P}_{k} 
+ \sum_{k=0}^{m_{0}} \Delta \widetilde{P}_{k}
\notag\\
& \qquad \qquad 
+ \sum_{k=1}^{m_{0}} \sum_{l=1}^{k} (-1)^{l} (l+1)
\langle \alpha, \alpha \rangle^{-1} C_{\alpha} 
y_{1}^{-l-2} \partial_{\eta_{1}}^{l} \widetilde{P}_{k-l} 
+ [P, S(y)] \widetilde{\enskip} 
 = 0
\notag\\
& \Leftrightarrow 
2 \langle \eta, \partial_{y} \rangle \widetilde{P}_{k+1} 
+ \Delta \widetilde{P}_{k}
\label{eq:commutative condition}
\\
& \qquad \qquad 
+ \sum_{l=1}^{k} (-1)^{l} (l+1) 
\langle \alpha, \alpha \rangle^{-1} C_{\alpha} 
y_{1}^{-l-2} \partial_{\eta_{1}}^{l} \widetilde{P}_{k-l} 
\notag\\
& \qquad \qquad 
+ (\mbox{the $(m_{0}-k)$-th order terms of } 
[P, S(y)]) \widetilde{\enskip} 
= 0 
\notag 
\end{align} 
for $k = 0, \dots, m_{0}$. 
Here, we have set $\widetilde{P}_{-1} = \widetilde{P}_{m_{0}+1} = 0$.

\begin{lemma} 
As a function of $y_{1}$, the order of the pole of
$\widetilde{P}_{k}$ at $y_{1} = 0$ is at most $k$. 
\end{lemma} 
\begin{proof} 
Denote by $O(F(y, \eta))$ the order of the pole
of a function $F(y, \eta)$ at $y_{1} = 0$. 
We prove this lemma by induction on $k$. 

By assumption, $\tilde{P}_{0}$ is constant in $y$. 
Therefore, $O(\widetilde{P}_{0}) = 0$. 
Now assume that $O(\widetilde{P}_{l}) \leq l$ 
for $l = 0, 1, \dots, k$. 
Then, $O(\Delta \widetilde{P}_{k})$ and 
$O(y_{1}^{-l-2} \partial_{\eta_{1}}^{l} \widetilde{P}_{k-l})$ 
are no greater than $k+2$. 
The $(m_{0}-k)$-th order terms of $[P, S(y)]$ come from
$[P_{l}, S(y)]$ ($l = 0, 1, \dots, k-1$). 
Because $S(y)$ is real analytic at $y_{1}=0$, 
$O([P_{l}, S(y)])$ is no greater than $k-1$, by the hypothesis of
induction. 
Hence $O(\langle \eta, \partial_{y}\rangle \widetilde{P}_{k+1})$ is 
no greater than $k+2$ by \eqref{eq:commutative condition}, 
and thus the lemma holds for $k+1$. 
\end{proof} 

Next, let $y' = (y_{2}, \dots, y_{n})$ and 
$\eta' = (\eta_{2}, \dots, \eta_{n})$, 
and let 
\begin{align*}
\widetilde{Q}_{k}(y', \eta_{1}, \eta')  
&= \lim_{y_{1} \rightarrow 0} y_{1}^{k} \widetilde{P}_{k} 
&
&
\mbox{and}
&
\widetilde{Q}_{k}'(y_{1}, y', \eta_{1}, \eta')  
&= \widetilde{P}_{k} - y_{1}^{-k} \widetilde{Q}_{k}. 
\end{align*}
Then, after substituting 
$\widetilde{P}_{k} 
= y_{1}^{-k} \widetilde{Q}_{k} + \widetilde{Q}_{k}'$ into
\eqref{eq:commutative condition}, 
and taking the limit 
$\lim_{y_{1} \rightarrow 0} (y_{1}^{k+2} \times 
\eqref{eq:commutative condition})$, 
we have 
\begin{equation}\label{eq:after rank one reduction}
-2 (k+1) \eta_{1} \widetilde{Q}_{k+1} 
+ k(k+1) \widetilde{Q}_{k} 
+ 
\sum_{l=1}^{k} 
(-1)^{l} (l+1) \langle \alpha, \alpha \rangle^{-1} C_{\alpha} 
\partial_{\eta_{1}}^{l} \widetilde{Q}_{k-l} 
= 0
\end{equation}
for $k = 0, 1, \dots, m_{0}$. 
This condition can be easily rephrased as follows.

\begin{lemma}\label{lemma:rank one reduction}
The polynomials $\widetilde{Q}_{k}$ ($k = 0, 1, \dots, m_{0} + 1$)
satisfy \eqref{eq:after rank one reduction} if and only if they
satisfy 
\begin{equation}\label{eq:one variable, rational}
\left[
-\frac{d^{2}}{d t^{2}} 
+ \frac{\langle \alpha, \alpha \rangle^{-1} C_{\alpha}}{t^{2}}, 
\sum_{k=0}^{m_{0}} t^{-k} 
\widetilde{Q}_{k} \left(y', \frac{d}{dt}, \eta' \right) 
\right] 
=0. 
\end{equation}
\end{lemma} 
With this lemma, the proof of 
Theorem~\ref{theorem:reflection invariance} is reduced to that for the
rank-one rational case.



\section{The rank-one rational case}
\label{section:The rank one rational case}
In this section, we solve 
\eqref{eq:one variable, rational} 
following Burchnall and Chaundy \cite{BC}. 

Let $L_{1}$ be a one variable Schr\"{o}dinger operator: 
\[
L_{1} 
= 
-\frac{d^{2}}{d t^{2}} + u(t).
\]
\begin{proposition}[\cite{BC}]
Assume that a differential operator $A_{m}$ of order $2m+1$ with
a constant principal symbol commutes
with $L_{1}$. 
Then 
$A_{m}$ can be expressed as 
\begin{equation*}
A_{m} 
= \sum_{k=0}^{m} 
\left( 
p_{k} \frac{d}{dt} - \frac{1}{2} p_{k}' 
\right) L_{1}^{m-k} 
\quad 
\mathrm{mod} \boldsymbol{C}[L_{1}], 
\end{equation*}
where $\{p_{j} ; j = 0, \dots, m+1\}$ is a solution 
of the system of functional equations 
\begin{equation}\label{eq:system of functional equation}
\left\{
\begin{split}
&
-\frac{1}{2} p_{j}''' 
+ 2 u p_{j}' 
- 2 p_{j+1}' 
+ u' p_{j} 
= 0 
\qquad 
(j = 0, 1, \dots, m), 
\\
& 
p_{0}' = 0, 
\\
& 
p_{m+1} = 0. 
\end{split}
\right.
\end{equation}
\end{proposition}
\begin{lemma}\label{lemma:rank one rational}
If the above operator $A_{m}$ commutes with 
\[L_{1} 
= 
-\frac{d^{2}}{dt^{2}} 
+ \frac{\langle \alpha, \alpha \rangle^{-1} C_{\alpha}}{t^{2}},
\] 
then there exists $k \in \{0, 1, \dots, m\}$ such that 
\[
C_{\alpha} = k(k+1) \langle \alpha, \alpha \rangle. 
\]
\end{lemma} 
\begin{proof}
First, we prove that the solution of 
\eqref{eq:system of functional equation} 
can be expressed as 
\begin{equation}\label{eq:solution of system of equations} 
p_{j} 
= 
\sum_{i=0}^{j} c_{j, i} t^{-2i}, 
\end{equation}
with suitable constants $c_{j, i}$, by induction on $j$.

Because \eqref{eq:system of functional equation} is linear in
$\{p_{j}\}$ and $p_{0}' = 0$, we may set $c_{0,0} = 1$. 
Suppose that $p_{0}, \dots, p_{j}$ are expressed as 
\eqref{eq:solution of system of equations}. 
Then \eqref{eq:system of functional equation} implies 
\begin{align*}
p_{j+1}' 
&= 
-\frac{1}{4} \sum_{i=0}^{j} c_{j, i} (-2i)(-2i-1)(-2i-2)
t^{-2i-3} 
\\
& \qquad 
+ \langle \alpha, \alpha \rangle^{-1} C_{\alpha} t^{-2} 
\sum_{i=0}^{j} c_{j, i} (-2i) t^{-2i-1} 
- \langle \alpha, \alpha \rangle^{-1} C_{\alpha} t^{-3} 
\sum_{i=0}^{j} c_{j, i} t^{-2i}
\\
&= 
\sum_{i=0}^{j} (2i+1) 
\{i(i+1) - \langle \alpha, \alpha \rangle^{-1} C_{\alpha}\} 
c_{j, i} t^{-2i-3}.
\end{align*}
Therefore, if we set 
\begin{align}
c_{j+1, i+1} 
&= \frac{2i+1}{2i+2} 
\{\langle \alpha, \alpha \rangle^{-1} C_{\alpha} - i(i+1)\} 
c_{j,i},
\label{eq:recursion}
\end{align}
with $c_{j+1,0}$ arbitrary for $j \geq 0$, 
then $p_{j+1}$ is also expressed as 
\eqref{eq:solution of system of equations}.

Now, by \eqref{eq:recursion}, 
we have 
\begin{align*}
c_{m+1, m+1} 
&= 
\frac{2m+1}{2m+2} 
\{\langle \alpha, \alpha \rangle^{-1} C_{\alpha} - m(m+1)\}
c_{m, m} 
\\
&= \prod_{k=0}^{m} 
\frac{2k+1}{2k+2} 
\{\langle \alpha, \alpha \rangle^{-1} C_{\alpha} - k(k+1)\}. 
\end{align*}
If $\{p_{j}\}$ is a solution of 
\eqref{eq:system of functional equation}, 
$c_{m+1, m+1}$ must be zero. 
Thus $C_{\alpha} = k(k+1) \langle \alpha, \alpha \rangle$ for some 
$k \in \{0, 1, \dots, m\}$. 
\end{proof} 

Now, we return to the proof of 
Theorem~\ref{theorem:reflection invariance}. 

Note that $\widetilde{P}_{0} = \widetilde{Q}_{0}$ because
$\widetilde{P}_{0}$ is constant in $y$. 
Then, because 
$C_{\alpha} \not= k(k+1) \langle \alpha, \alpha \rangle$ for any
$k \in \boldsymbol{Z}$, 
Lemma~\ref{lemma:rank one rational} and 
Lemma~\ref{lemma:rank one reduction} imply that 
$\widetilde{P}_{0} = \widetilde{Q}_{0}$ is even in $\eta_{1}$; 
that is, it can be expressed as 
$\widetilde{P}_{0} 
= \sum_{k = 0}^{[m_{0} / 2]} 
\eta_{1}^{2k} \widetilde{P}_{0, k} (\eta')$. 
Moreover, $\widetilde{P}_{0, k} (\eta')$ is also 
$r_{\alpha}$-invariant, because 
$\eta_{2}, \dots, \eta_{n}$ are the symbols of 
directional differentials along $H_{\alpha}$. 
Therefore, $\widetilde{P}_{0}$ is $r_{\alpha}$-invariant. 
{\hspace*{\fill}$\square$\par\bigskip}



\section{Hyperplane arrangement in the generic case} 
\label{section:Hyperplane arrangement in a generic case} 
In this section, we address the problem of determining the permissible
kinds of hyperplane arrangements when the potential function is
generic, i.e. in the case that 
$C_{\alpha} 
\not= 
k(k+1) 
\langle \alpha, \alpha \rangle$ 
for any integer $k$ and any $\alpha \in \mathcal{H}$. 
In order to exclude trivial cases, 
we assume that the principal symbol of $P$ is not a polynomial in
$\sum_{i=1}^{n} \xi_{i}^{2}$. 
Moreover, in order to avoid the possibility of reduction to a
lower-dimensional case, 
we assume the ``irreducibility'' of the hyperplane
arrangement $\mathcal{H}$, as defined below. 
\begin{definition}\label{def:irreducibility of H} 
A finite subset $\mathcal{H}$ of mutually non-parallel vectors in
$\boldsymbol{R}^{n}$ is {\it irreducible} if it satisfies the following
conditions: 
\begin{enumerate} 
\item[(I1)]
$\mathcal{H}$ spans $\boldsymbol{R}^{n}$ as an $\boldsymbol{R}$-vector
space. 
\item[(I2)]
$\mathcal{H}$ cannot be partitioned into the union of two
proper subsets such that each vector in one subset is
orthogonal to each vector in the other. 
\end{enumerate} 
\end{definition} 

Let $W$ be the reflection group generated by 
$\{r_{\alpha} ; \alpha \in \mathcal{H}\}$ and $\overline{W}$
be the closure of $W$ in $O(n)$. 
\begin{proposition} 
If $W$ is an infinite group, then $\overline{W}$ is
isomorphic to $O(n)$. 
\end{proposition}

\begin{proof} 
By a general theory of topological groups, $\overline{W}$ is
a closed subgroup of $O(n)$, or in other words, 
a compact Lie subgroup. 

Because $\# W = \infty$, 
$\overline{W}$ contains a subgroup $T$ isomorphic to $SO(2)$. 
Let 
$V^{T} 
= \{v \in \boldsymbol{R}^{n} ; t v = v, \enskip \forall t \in T\}$. 
If $\mathcal{H} \subset V^{T}$, then $T$ acts trivially
on $\boldsymbol{R}^{n}$, by (I1). 
This contradicts the relation $T \simeq SO(2)\subset O(n)$. 
Therefore, there exists $\alpha \in \mathcal{H}$ such that 
$t \alpha \not= \alpha$ for any $t \in T$ sufficiently close to $e$. 
We can choose $t \in T$ such that 
the closure of 
$\langle r_{\alpha}, t r_{\alpha} t^{-1} = r_{t \alpha} \rangle$ is
isomorphic to $O(2)$. 
Let $\alpha_{1} = \alpha$, $\alpha_{2} = t \alpha$, 
$V_{2} = \boldsymbol{R} \alpha_{1} + \boldsymbol{R} \alpha_{2}$, 
$U_{2} = V_{2}^{\bot}$ and 
$G_{2} 
= 
\overline{\langle r_{\alpha_{1}}, r_{\alpha_{2}} \rangle}$. 
Then, obviously, $\boldsymbol{R}^{n} = V_{2} \oplus U_{2}$, 
and $G_{2}$ is a closed subgroup of $\overline{W}$. 
Because $G_{2}$ acts trivially on $U_{2}$, 
we have 
$G_{2} \simeq 
\begin{pmatrix} 
O(2) & O 
\\
O & I_{n-2} 
\end{pmatrix} 
\hookrightarrow 
O(n)$. 

Now, let us define a $k$-dimensional subspace $V_{k}$, the orthogonal
complement $U_{k}$ of $V_{k}$, and a compact subgroup $G_{k}$ of 
$\overline{W}$ inductively as follows. 
For $k<n$, not all vectors in $\mathcal{H}$ are contained in
$V_{k}$, by (I1). 
Therefore, 
$\mathcal{H}_{k} 
:= \mathcal{H} \setminus (\mathcal{H} \cap V_{k})$ 
is not empty. 
Since $\mathcal{H}_{k} \not\subset U_{k}$, by (I2), 
we can choose a vector $\alpha_{k+1} \in \mathcal{H}_{k}$ 
satisfying $\alpha_{k+1} \not\in U_{k}$. 
Then, let $V_{k+1} = \boldsymbol{R} \alpha_{k+1} + V_{k}$, 
$U_{k+1} = \alpha_{k+1}^{\bot} \cap U_{k} = V_{k+1}^{\bot}$, 
and $G_{k}$ be the closure of
the group generated by $G_{k}$ and $r_{\alpha_{k+1}}$. 
Clearly, $V_{k+1}$ is a $G_{k+1}$-invariant subspace, 
and $G_{k+1}$ acts trivially on $U_{k+1}$. 

Next, choose an orthonormal basis $\{f_{1}, \dots, f_{n}\}$ of
$\boldsymbol{R}^{n}$ such that $\{f_{1}, \dots, f_{k}\}$ is a basis of
$V_{k}$. 
By induction on $k$, we now show that the realization of $G_{k}$ with
respect to this basis is 
$\begin{pmatrix} 
O(k) & O 
\\
O & I_{n-k} 
\end{pmatrix}$. 

The case $k=2$ has been demonstrated. 

Now, assume $G_{k}$ to be realized as above. 
Then, denote by 
$\boldsymbol{a} 
= 
\begin{pmatrix} 
\boldsymbol{a'}
\\
a_{k+1} 
\\
\boldsymbol{0}
\end{pmatrix}$  
($\boldsymbol{a'} \in \boldsymbol{R}^{k}$) 
the coordinates of $\alpha_{k+1}$ with respect to $\{f_{i}\}$. 
Note that $\boldsymbol{a'} \not= \boldsymbol{0}$ and
$a_{k+1} \not= 0$, 
because $\alpha_{k+1} \not\in V_{k+1} \cap U_{k}$ and
$\alpha_{k+1} \not\in V_{k}$. 
By the assumption of induction, 
the Lie algebra of $G_{k}$ is realized as 
$\begin{pmatrix} 
\mathfrak{o}(k) & O 
\\
O & O 
\end{pmatrix}$. 
Let 
$X = 
\begin{pmatrix} 
X' & O 
\\
O & O 
\end{pmatrix}$ 
($X' \in \mathfrak{o}(k)$) be an element of $\mathrm{Lie} (G_{k})$. 
Because the representation matrix of $r_{\alpha_{k+1}}$ is 
$I - 2|\boldsymbol{a}|^{-2} 
\boldsymbol{a} {}^{t}\boldsymbol{a}$, 
we have 
\begin{align*}
\mathrm{Ad}(r_{\alpha_{k+1}}) X 
&= 
(I - 2|\boldsymbol{a}|^{-2} 
\boldsymbol{a} {}^{t}\boldsymbol{a}) 
X 
(I - 2|\boldsymbol{a}|^{-2} 
\boldsymbol{a} {}^{t}\boldsymbol{a}) 
\\
&= 
\begin{pmatrix} 
X' 
- 2 |\boldsymbol{a}|^{-2} 
(\boldsymbol{a'} {}^{t}\boldsymbol{a'} X' 
+ X' \boldsymbol{a'} {}^{t}\boldsymbol{a'}) 
& 
- 2 |\boldsymbol{a}|^{-2} a_{k+1} X' 
{}^{t}\boldsymbol{a'}  
& O 
\\
- 2 |\boldsymbol{a}|^{-2} a_{k+1} 
{}^{t}\boldsymbol{a'} X'
& 0 & O 
\\
O & O & O 
\end{pmatrix}. 
\end{align*}
Here, we have used 
${}^{t}\boldsymbol{a'} X' \boldsymbol{a'} = 0$, 
as ${}^{t}\boldsymbol{a'} X' \boldsymbol{a'}$ 
is a $1 \times 1$-alternative matrix. 
Since $\boldsymbol{a'} \not= \boldsymbol{0}$ and 
$a_{k+1} \not=0$, there exists an $X' \in \mathfrak{o}(k)$ such that 
\[\begin{pmatrix} 
O
& 
- 2 |\boldsymbol{a}|^{-2} a_{k+1} X' 
{}^{t}\boldsymbol{a'}  
\\
- 2 |\boldsymbol{a}|^{-2} a_{k+1} 
{}^{t}\boldsymbol{a'} X'
& 0 
\end{pmatrix} 
\not= O. 
\]
As a Lie algebra, $\mathfrak{o}(k+1)$ is generated by this
matrix and $\mathfrak{o}(k)$. 
Therefore, 
$\mathrm{Lie} (G_{k+1})$ is realized as  
$\begin{pmatrix} 
\mathfrak{o}(k+1) & O 
\\
O & O 
\end{pmatrix}$, 
and $G_{k+1}$ as 
$\begin{pmatrix} 
O(k+1) & O
\\
O & I_{n-k-1}
\end{pmatrix}$, 
because $G_{k+1}$ acts trivially on $U_{k+1}$. 

Finally, 
$O(n) = G_{n} \subset \overline{W}$ implies 
$\overline{W} = O(n)$. 
\end{proof}

\begin{corollary}\label{cor:O(n)-inv} 
If $\mathcal{H}$ is irreducible and $W$ is an infinite
group, then any $W$-invariant polynomial 
in $\boldsymbol{C}[\boldsymbol{R}^{n}]$ is a
polynomial in $\sum_{i=1}^{n} \xi_{i}^{2}$. 
\end{corollary} 

\begin{theorem}\label{thm:main in section 4} 
Suppose that the principal symbol of $P$ is constant in $x$ 
and is not a polynomial in $\sum_{i=1}^{n} \xi_{i}^{2}$. 
Then, if $R(x)$ in \eqref{eq:Schrodinger} is generic and $\mathcal{H}$
is irreducible, 
either $W$ is a finite reflection group or $[L, P] \not= 0$. 
\end{theorem} 
\begin{proof} 
Assume that $L$ and $P$ are commutative. 
Then, by Theorem~\ref{theorem:reflection invariance}, 
the principal symbol $\widetilde{P}_{0}$ of $P$ is a $W$-invariant
polynomial. 
If $W$ is infinite, $\widetilde{P}_{0}$ must be a polynomial in 
$\sum_{i=1}^{n} \xi_{i}^{2}$, by Corollary~\ref{cor:O(n)-inv}. 
However, this contradicts the assumption. 
Therefore $W$ is a finite reflection group. 
\end{proof}

By this theorem, in generic cases, we need consider only the case in
which $\mathcal{H}$ is a subset of the root system of a finite
reflection group. 



\section{Determination of the potential -- general situation} 
\label{section:Determination of the potential -- general
setting}

Assume that $\mathcal{H}$ is irreducible and 
that $R(x)$ is generic. 
Then, as stated above, 
we need consider only the case in which 
$W = \langle r_{\alpha} ; \alpha \in \mathcal{H} \rangle$ 
is a finite reflection group; 
that is, we may regard $\mathcal{H}$ as a subset of the positive root
system $\Sigma^{+}$ of $W$. 
In subsequent sections, we determine the potential function $R(x)$
in the cases that the root system $\Sigma$ is of
type $A$, $B$ and $D$ under some conditions. 
In this section, we explain the general situation.

Let $P$ be a commutant of $L$ with a constant principal symbol. 
Because $R(x)$ is generic, the principal symbol of $P$ is
$W$-invariant, 
by Theorem~\ref{theorem:reflection invariance}. 
We assume the following conditions: 
\begin{enumerate} 
\item
$P$ is real analytic in the domain where $L$ is defined. 
\item
The order of $P$ is the smallest degree of $W$ larger than
$2$. 
\end{enumerate} 

In general, for a differential operator 
$D = \sum_{p} a_{p} (x) \partial^{p}$, we define ${}^{t}D$ as 
\[
{}^{t} D 
= \sum_{p} (-1)^{|p|} \partial^{p} \circ a_{p} (x), 
\]
and call it the adjoint operator of $P$. 
Because $L$ is self-adjoint (i.e. ${}^{t} L = L$), 
if $P$ commutes with $L$, so does ${}^{t} P$. 
Therefore, 
we may assume that $P$ is (skew-) self-adjoint, i.e. 
${}^{t} P = (-1)^{\mathrm{ord} P} P$.



\section{Determination of the potential -- type $A_{n-1}$} 
\label{section:Determination of the potential -- type $A_{n-1}$} 

The arguments hereafter are quite similar to those in \cite{OS}. 
There, the Weyl group invariance of $L$ and $P$ is assumed, 
but here this assumption is not made. 
This is the most important difference between the situations
considered here and in that work.

The root system of type $A_{n-1}$ is realized in the hyperplane 
\[
V = 
\left\{
(x_{1}, \dots, x_{n}) \in \boldsymbol{R}^{n} ; 
\sum_{i=1}^{n} x_{i} = 0
\right\},
\]
and we choose a positive system as 
\[
\Sigma^{+} = \{e_{i} - e_{j} ; 1 \leq i < j \leq n\}.
\]
By virtue of this realization, 
the Schr\"{o}dinger operator \eqref{eq:Schrodinger} is extended to the
operator 
\begin{align}
L &= 
- \Delta 
+ R(x), 
&
R(x) 
&= 
\sum_{1 \leq i < j \leq n} 
\frac{C_{ij}}{(x_{i} - x_{j})^{2}} 
+ 
\widetilde{R}(x), 
\label{eq:A-01}
\end{align}
defined on some open subset of $\boldsymbol{R}^{n}$, 
where $\widetilde{R}(x)$ is a real analytic at
$x=0$ and $L$ commutes with 
\[
\Delta_{1} 
= 
\sum_{i=1}^{n} \partial_{i}. 
\]
Note that some of the constants $C_{ij}$ may be zero, 
because $\mathcal{H}$ might not coincide with $\Sigma^{+}$. 

As a commutant $P$ of $L$, 
we can choose 
\begin{equation}\label{eq:A-02} 
P = 
\sum_{1 \leq i < j < k \leq n} 
\partial_{i} \partial_{j} \partial_{k} 
+ 
\sum_{i=1}^{n} a_{1}^{i} \partial_{i} 
+ a_{0},
\end{equation}
which commutes with $\Delta_{1}$. 

As seen from Remark~2.4 of  \cite{OS}, 
the equations $[L,P]=0$, $[L, \Delta_{1}]=0$ and $[\Delta_{1}, P]=0$
imply that $R(x)$ can be expressed as 
\begin{equation}\label{eq:A-1} 
R(x) = 
\sum_{1 \leq i < j \leq n} u_{ij}(x_{i} - x_{j}) 
\end{equation} 
with suitable functions 
$u_{ij}(t) = C_{ij} t^{-2} + \gamma_{ij}(t)$, 
where $\gamma_{ij}(t)$ is real analytic at $t = 0$.  
For convenience, let $u_{ij}(t) = u_{ji}(-t)$ for $j < i$. 
\begin{lemma} 
We can choose $a_{1}^{i}$ as 
\begin{equation}\label{eq:A-1.5}
a_{1}^{i} 
= \frac{1}{2} 
\sum_{j<k \atop \not= i} 
u_{jk}(x_{j} - x_{k}), 
\end{equation}
because we are free to choose $u_{ij}$ appropriately. 
\end{lemma}
\begin{proof} 
The second-order terms of $[L, P] = 0$ imply 
\begin{align}
\partial_{i}^{2} & : \partial_{i} a_{1}^{i} = 0, 
\label{eq:A-2}\\
\partial_{i} \partial_{j} 
& : \partial_{j} a_{1}^{i} + \partial_{i} a_{1}^{j} 
= -\frac{1}{2} \sum_{k \not= i, j} \partial_{k} R 
= \frac{1}{2} (\partial_{i} + \partial_{j}) R. 
\label{eq:A-3}
\end{align}
Also, by \eqref{eq:A-3}, we have 
\begin{align}
& \partial_{j} \partial_{k} a_{1}^{i} 
+ \partial_{i} \partial_{k} a_{1}^{j} 
= \frac{1}{2} \partial_{k} (\partial_{i}+\partial_{j}) R, 
\label{eq:A-4}
\\
& \partial_{k} \partial_{i} a_{1}^{j} 
+ \partial_{j} \partial_{i} a_{1}^{k} 
= \frac{1}{2} \partial_{i} (\partial_{j}+\partial_{k}) R, 
\label{eq:A-5}
\\
& \partial_{i} \partial_{j} a_{1}^{k} 
+ \partial_{k} \partial_{j} a_{1}^{i} 
= \frac{1}{2} \partial_{j} (\partial_{k}+\partial_{i}) R.  
\label{eq:A-6}
\end{align}
Taken together, \eqref{eq:A-4}, \eqref{eq:A-6} and \eqref{eq:A-5}
imply 
\begin{equation*} 
\partial_{j} \partial_{k} a_{1}^{i} 
= \frac{1}{2} \partial_{j} \partial_{k} R 
= - \frac{1}{2} u_{jk}''(x_{j} - x_{k}). 
\end{equation*} 
Moreover, because the relation $[\Delta_{1}, P] = 0$ implies
$\Delta_{1} a_{1}^{i} = 0$, we have the following: 
\[
\partial_{j}^{2} a_{1}^{i} 
= \partial_{j} (\partial_{j} - \Delta_{1}) a_{1}^{i} 
= - \sum_{k \not= i, j} \partial_{j} \partial_{k} a_{1}^{i} 
= \frac{1}{2} \sum_{k \not= i, j} u_{jk}''(x_{j} - x_{k}).
\] 
Therefore, we may put 
\begin{equation*}
a_{1}^{i} 
= 
\frac{1}{2} \sum_{j < k \atop \not= i} 
u_{jk}(x_{j} - x_{k}) 
+ \sum_{j \not= i} p_{i,j} x_{j} + q_{i}. 
\end{equation*}
Equation \eqref{eq:A-3} implies $p_{i,j} = - p_{j,i}$, 
and $\Delta_{1} a_{1}^{i} = 0$ implies $\sum_{j \not= i} p_{i,j} = 0$. 
Next, let $\overset{\circ}{q} = (\sum_{i=1}^{n} q_{i}) / n$, 
$\tilde{q}_{i} = q_{i} - \overset{\circ}{q}$ and 
\[
\tilde{u}_{ij}(t) 
= u_{ij}(t) + 2 p_{i,j} t + \beta_{ij} 
\] 
for $1 \leq i < j \leq n$,
where the quantities $\beta_{ij}$ are given by 
\begin{align*}
& \beta_{i1} = \beta_{1i} = 
-2 \tilde{q}_{i} 
\qquad (i = 2,\dots, n-2, \mbox{ if $n > 3$}), 
\\
& \beta_{n-1,i} = \beta_{1,n-1} = 
-2 (\tilde{q}_{1} + \tilde{q}_{n-1}),
\\
& \beta_{n,1} = \beta_{1,n} = 
-2 (\tilde{q}_{1} + \tilde{q}_{n}), 
\\
& \beta_{n,n-1} = \beta_{n-1,n} 
= 
2 \tilde{q}_{1}, \quad \mbox{and}
\\
& \beta_{ij} = 0 \qquad (\mbox{otherwise}).  
\end{align*} 
Then, because $\sum_{i=1}^{n} \tilde{q}_{i} = 0$, 
we have 
$\sum_{i<j} \beta_{ij} 
= -2 \sum_{i=2}^{n-2} \tilde{q}_{i} 
-2 (\tilde{q}_{1} + \tilde{q}_{n-1})
-2 (\tilde{q}_{1} + \tilde{q}_{n})
+ 2 \tilde{q}_{1} 
= 0$ and  
$\sum_{j<k \atop \not= i} \beta_{ij} = 2 \tilde{q}_{i}$. 
Therefore, 
\begin{align*}
\sum_{1 \leq i < j \leq n} u_{ij}(x_{i} - x_{j}) 
&= 
\sum_{1 \leq i < j \leq n} 
(\tilde{u}_{ij}(x_{i} - x_{j}) - 2 p_{i,j} (x_{i} - x_{j}) 
- \beta_{ij}) 
\\
&= 
\sum_{1 \leq i < j \leq n} 
\tilde{u}_{ij}(x_{i} - x_{j}) 
\end{align*}
and 
\begin{align*}
a_{1}^{i} 
&= 
\frac{1}{2} \sum_{j < k \atop \not= i} 
(\tilde{u}_{jk}(x_{j} - x_{k}) - 2 p_{j,k} (x_{j} - x_{k}) 
- \beta_{jk})
+ \sum_{j \not= i} p_{i,j} x_{j} + \tilde{q}_{i} 
+ \overset{\circ}{q}
\\
&= 
\frac{1}{2} \sum_{j < k \atop \not= i} 
\tilde{u}_{jk}(x_{j} - x_{k}) 
+ \overset{\circ}{q}. 
\end{align*}
Hence, by subtracting $\overset{\circ}{q} \Delta_{1}$ from $P$, 
we obtain \eqref{eq:A-1.5}. 
\end{proof} 

The condition ${}^{t}P = -P$ is equivalent to 
\begin{equation*} 
a_{0} 
= \frac{1}{2} \sum_{i=1}^{n} \partial_{i} a_{1}^{i} = 0. 
\end{equation*} 
Therefore, the zeroth-order term of the relation $[L, P] = 0$ implies 
\[
\sum_{i<j<k} \partial_{i} \partial_{j} \partial_{k} R 
+ \sum_{i} a_{1}^{i} \partial_{i} R = 0. 
\]
Applying \eqref{eq:A-1} and \eqref{eq:A-1.5} to this equation, 
we have 
\begin{equation}\label{eq:A-7}
\sum_{i<j} 
\left( 
\sum_{p \not= i,j} 
(u_{pj}(x_{p} - x_{j}) - u_{pi}(x_{p} - x_{i})) 
\right) 
u_{ij}'(x_{i} - x_{j}) 
= 0. 
\end{equation}
Then, because 
$u_{ij}(t) - C_{ij} t^{-2}$ is real analytic at $t=0$, 
$\lim_{x_{j} \rightarrow x_{i}} 
((x_{i} - x_{j})^{3} \times \eqref{eq:A-7})$ gives 
\begin{equation}\label{eq:A-8}
C_{ij} 
\sum_{p \not= i, j} 
(u_{pi}(x_{p} - x_{i}) - u_{pj}(x_{p} - x_{i})) 
 = 0. 
\end{equation}
Moreover, $\lim_{x_{k} \rightarrow x_{i}} 
((x_{k} - x_{i})^{2} \times \eqref{eq:A-8})$ gives  
\begin{equation}\label{eq:A-9}  
C_{ij} (C_{ki} - C_{kj}) = 0 
\end{equation}
for $k \not= i, j$. 
Because $\mathcal{H}$ is not empty, 
there exist $i_{1}$ and $i_{2}$ ($i_{1} \not= i_{2}$) such that 
$C_{i_{1} i_{2}} \not= 0$. 
Then, employing an appropriate coordinate transformation, 
we can put $i_{1} = 1$ and $i_{2} = 2$. 
Therefore, by \eqref{eq:A-9}, 
we have $C_{1i} = C_{2i}$ for $i \geq 3$. 
The condition (I2) and the relation $C_{12} \not= 0$ imply that there
exists $i_{3}$ such that $C_{1 i_{3}} = C_{2 i_{3}} \not= 0$. 
Again, our ability to apply coordinate transformations allows us to
choose $i_{3} = 3$. 
Then, from \eqref{eq:A-9}, 
we find $C_{12} = C_{23} = C_{13}$ 
and $C_{1i} = C_{3i} = C_{2i}$ for $i \geq 4$. 
In the same way, 
we can show inductively that $C_{ij}$ depends on neither $i$
nor $j$. 
In particular, none of them are zero. 

The fact that $C_{ij} \not= 0$ and equation \eqref{eq:A-8} together
imply 
\begin{equation}\label{eq:A-10} 
u_{ki}'(t) = u_{kj}'(t). 
\end{equation} 
Then, because $u_{ij}(t) = u_{ji}(-t)$, 
\eqref{eq:A-10} implies $u_{ik}'(t) = u_{jk}'(t)$, 
and we have 
\begin{equation*}
u_{ij}'(t) = u_{ik}'(t) = u_{jk}'(t) = -u_{kj}'(-t) 
= - u_{ij}'(-t). 
\end{equation*}
Therefore $u_{ij}(t)$ is an even function 
and, by \eqref{eq:A-10}, 
there exist constants $c_{ij}$ ($1 \leq i < j \leq n$) and 
an even function $u(t)$ such that 
\begin{equation*} 
u_{ij}(t) = u(t) + c_{ij}. 
\end{equation*} 
Because $u(t)$ is fixed only up to an arbitrary constant, 
we can choose the $c_{ij}$ so that $\sum_{i<j} c_{ij} = 0$. 
From \eqref{eq:A-8}, 
we obtain 
$\sum_{p \not= i,j} c_{pi} = \sum_{p \not= i,j} c_{pj} \Leftrightarrow
\sum_{p \not= i} c_{pi} = \sum_{p \not= j} c_{pj}$. 
This means that $\widetilde{c} = \sum_{p \not= i} c_{pi}$ does not
depend on $i$. 
Then, because $\sum_{i<j} c_{ij} = 0$, we have 
\begin{align*}
R(x) 
= 
\sum_{1 \leq i < j \leq n} 
u_{ij}(x_{i} - x_{j}) 
= 
\sum_{1 \leq i < j \leq n} 
(u(x_{i} - x_{j}) + c_{ij}) 
= 
\sum_{1 \leq i < j \leq n} 
u(x_{i} - x_{j}) 
\end{align*}
and 
\begin{align*}
a_{1}^{i} 
= 
\frac{1}{2} \sum_{j < k \atop \not= i} 
u_{jk}(x_{j} - x_{k}) 
= 
\frac{1}{2} \sum_{j < k \atop \not= i} 
(u(x_{j} - x_{k}) + c_{jk}) 
= 
\frac{1}{2} \sum_{j < k \atop \not= i} 
u(x_{j} - x_{k}) 
- \frac{\tilde{c}}{2}. 
\end{align*}
Then, the freedom we have to add $(\widetilde{c}/2) \Delta_{1}$ to $P$
allows us to realize the condition $c_{ij} = 0$ for all $i \not= j$. 
In this case, \eqref{eq:A-7} becomes 
\begin{equation*}
\sum_{i<j} 
\left( 
\sum_{p \not= i,j} 
(u(x_{p} - x_{j}) - u(x_{p} - x_{i})) 
\right) 
u'(x_{i} - x_{j}) 
= 0. 
\end{equation*}
In \cite{OS}, Oshima and Sekiguchi solved this functional
equation. 
They obtained the solution 
\[
u(t) = c_{1} \wp(t| 2\omega_{1}, 2\omega_{2}) + c_{2}, 
\]
where $c_{1}$ and $c_{2}$ are arbitrary constants and 
$\wp(t| 2\omega_{1}, 2\omega_{2})$ is the Weierstrass elliptic
function with primitive periods $2\omega_{1}$ and
$2\omega_{2}$.

Combining the above results, we have proved the following theorem. 
\begin{theorem}\label{thm:A-type last} 
If $L$ in \eqref{eq:A-01} commutes with $P$ in
\eqref{eq:A-02}, 
then there exist constants $c_{1}$ and $c_{2}$ such that 
\[
R(x) 
= c_{1} 
\sum_{1 \leq i < j \leq n} 
\wp(x_{i} - x_{j}| 2\omega_{1}, 2\omega_{2}) 
+ c_{2}.
\]
\end{theorem}



\section{Determination of the potential 
-- types $B_{n}$ and $D_{n}$} 
\label{section:Determination of the potential 
-- type $B_{n}$ and $D_{n}$} 

Assume $W$ to be 
of type $B_{n}$ ($n \geq 2$) or $D_{n}$ ($n \geq 4$). 
The root systems of type $B_{n}$ and $D_{n}$ are realized in
$\boldsymbol{R}^{n}$. 
We choose 
\begin{align*}
\Sigma^{+} 
&= 
\{e_{i} \pm e_{j}; 1 \leq i < j \leq n\} 
\cup \{e_{i} ; 1 \leq i \leq n\} 
\qquad 
\mbox{for $B_{n}$-type and} 
\\
\Sigma^{+} 
&= 
\{e_{i} \pm e_{j}; 1 \leq i < j \leq n\} 
\qquad 
\mbox{for $D_{n}$-type} 
\end{align*}
as their positive systems. 
In these cases, the Schr\"{o}dinger operator \eqref{eq:Schrodinger} is 
\begin{align}
L &= 
- \Delta + R(x), 
\notag
\\
R(x) &= 
\sum_{1 \leq i < j \leq n} 
\left(\frac{C_{ij}^{+}}{(x_{i} + x_{j})^{2}} 
+ \frac{C_{ij}^{-}}{(x_{i} - x_{j})^{2}}
\right)
+ \sum_{i=1}^{n} \frac{C_{i}}{x_{i}^{2}} 
+ \tilde{R}(x), 
\label{eq:BD-2}
\end{align} 
where $\tilde{R}(x)$ is real analytic at $x=0$, 
and $C_{i} = 0$ for $i = 1, \dots, n$ in the 
$D_{n}$ case. 
 
As the commutant $P$ satisfying the two conditions in 
\S\ref{section:Determination of the potential -- general
setting}, 
we can choose 
\begin{equation}\label{eq:BD-3}
P = 
\sum_{1 \leq i < j \leq n} \partial_{i}^{2} \partial_{j}^{2} 
+ \sum_{i=1}^{n} a_{2}^{i} \partial_{i}^{2} 
+ \sum_{1 \leq i < j \leq n} a_{11}^{ij} \partial_{i} \partial_{j} 
+ \sum_{i=1}^{n} a_{1}^{i} \partial_{i} 
+ a_{0}.
\end{equation}
For convenience, we set $a_{11}^{ij} = a_{11}^{ji}$ for $j < i$. 
\begin{remark}\label{rem:D_4} 
In the $D_{4}$ case, other choices of $P$ are possible, 
since 
\[
P 
= 
c_{1} 
\sum_{1 \leq i < j \leq 4} 
\partial_{i}^{2} \partial_{j}^{2} 
+ c_{2} 
\partial_{1} \partial_{2} \partial_{3} \partial_{4} 
+ 
\mbox{(lower-order terms)} 
\]
satisfies the two conditions in 
\S\ref{section:Determination of the potential -- general
setting} for any $c_{1}$ and $c_{2}$. 
If $c_{1} = 1$ and $c_{2} = \pm 6$, 
the fourth order term of $P$ changes to 
\[
\frac{3}{4}\sum_{i=1}^{4} \partial_{y_{i}^{\pm}}^{4} 
-\frac{1}{2} \sum_{1 \leq i < j \leq 4} 
\partial_{y_{i}^{\pm}}^{2} \partial_{y_{j}^{\pm}}^{2} 
\]
through the orthogonal coordinate transformation defined by 
\begin{equation*}
\begin{pmatrix}
x_{1} \\ x_{2} \\ x_{3} \\ x_{4} 
\end{pmatrix} 
\mapsto 
\begin{pmatrix}
y_{1} \\ y_{2} \\ y_{3} \\ y_{4} 
\end{pmatrix} 
:= 
\frac{1}{2} 
\begin{pmatrix} 
1 & \pm 1 & \pm 1 & \pm 1 
\\
\pm 1 & 1 & -1 & -1 
\\
\pm 1 & -1 & 1 & -1 
\\
\pm 1 & -1 & -1 & 1 
\end{pmatrix} 
\begin{pmatrix} 
x_{1} \\ x_{2} \\ x_{3} \\ x_{4} 
\end{pmatrix}. 
\end{equation*}
%
%
%
%
Therefore, the situation regarding the commutator of this operator and
$L$ is equivalent to that for the operator in \eqref{eq:BD-3} and $L$. 
In the case $c_{2} \not= \pm 6 c_{1}$, 
it can be shown that $R(x)$ can be expressed as 
\begin{equation}\label{eq:twisted D_4}
\sum_{1 \leq i < j \leq 4} (u(x_{i} - x_{j}) + u(x_{1} + x_{j})) 
+ 
\sum_{i=1}^{4} v(x_{i}), 
\end{equation}
with $u(t)$ and $v(t)$ appropriately chosen even functions. 
This is identical to the assertion of 
Theorem~\ref{theorem:BD-final} below. 
\end{remark}

Now, let us return to the situation described prior to
Remark~\ref{rem:D_4}. 
The third-order terms of the relation $[L, P] = 0$ imply 
\begin{align}
& \partial_{i}^{3} :
& & \hspace{-30mm} 
\partial_{i} a_{2}^{i} = 0, 
\label{eq:BD-4}\\
& \partial_{i}^{2} \partial_{j} : 
& & \hspace{-30mm} 
\partial_{j} a_{2}^{i} + \partial_{i} a_{11}^{ij} = -\partial_{j} R, 
\label{eq:BD-5}\\
& \partial_{i} \partial_{j} \partial_{k} : 
& & \hspace{-30mm} 
\partial_{k} a_{11}^{ij} + \partial_{i} a_{11}^{jk} 
+ \partial_{j} a_{11}^{ik} 
= 0. 
\label{eq:BD-6}
\end{align}
\begin{lemma}(\cite[Lemma~2.5]{OS})
\label{lemma:BD-1}
\begin{enumerate}
\item
Let $n \geq 3$. 
If the functions $u_{i}(x)$ ($1 \leq i \leq n$) and
$u_{ij}(x) = u_{ji}(x)$ ($1 \leq i < j \leq n$) of 
$x = (x_{1}, \dots, x_{n})$ 
satisfy 
\begin{equation}\label{eq:BD-7}
\partial_{j} u_{i} + \partial_{i} u_{ij} = 0
\qquad \mbox{and} \qquad 
\partial_{k} u_{ij} + \partial_{i} u_{jk} + \partial_{j} u_{ik} = 0, 
\end{equation}
then 
\[
\partial_{j}^{2} \partial_{k} u_{i} = 0 
\qquad \mbox{and} \qquad
\partial_{j} \partial_{k} \partial_{l} u_{i} = 0. 
\]
\item
Moreover, if they also satisfy 
\begin{equation}\label{eq:BD-8}
\partial_{i} u_{i} = 0
\end{equation} 
for $1 \leq i \leq n$, 
then
\begin{equation}\label{eq:BD-10}
\partial_{j}^{2} u_{ij} = 0, 
\qquad \mbox{and} \qquad 
\partial^{\alpha} u_{i} = \partial^{\alpha} u_{ij} = 0 
\qquad 
\mbox{if $|\alpha| \geq 3$}. 
\end{equation}
\item
If $n = 2$, the first relation in \eqref{eq:BD-7} and the relation
\eqref{eq:BD-8} imply \eqref{eq:BD-10}. 
\end{enumerate}
\end{lemma}

First, assume that $n \geq 3$, and let 
$u_{i} = a_{2}^{i} + R$ and $u_{ij} = a_{11}^{ij}$. 
Then, the relations \eqref{eq:BD-5} and \eqref{eq:BD-6} imply that
$u_{i}$ and $u_{ij}$ satisfy the conditions in
Lemma~\ref{lemma:BD-1}~(1). 
Therefore, 
$\partial_{j}^{2} \partial_{k} (a_{2}^{i} + R) 
= 0$ and $\partial_{j} \partial_{k} \partial_{l} (a_{2}^{i} + R) = 0$, 
which imply 
$\partial_{i} \partial_{j}^{2} \partial_{k} R 
= \partial_{i} \partial_{j} \partial_{k} \partial_{l} R 
= 0$. 
Therefore, $\partial_{i} \partial_{j} \partial_{k} R$ is a constant, 
and $\partial_{i} \partial_{j} R$ can be
expressed as 
\begin{equation*}
\partial_{i} \partial_{j} R 
= \sum_{k \not= i, j} c_{ijk} x_{k} 
+ \phi_{ij}(x_{i}, x_{j}),
\end{equation*}
for an appropriate choice of the constants $c_{ijk}$ and the function
$\phi_{ij}(x_{i}, x_{j})$. 
Note that this expression is also valid for the $B_{2}$ case, 
in which the first term of it is ignored. 
Now, from the relations \eqref{eq:BD-4} and \eqref{eq:BD-5}, 
we have 
$\partial_{i} \partial_{j} (\partial_{i}^{2} - \partial_{j}^{2}) R 
= \partial_{i} \partial_{j} 
\{\partial_{i}^{2} (R + a_{2}^{j}) - \partial_{j}^{2} (R + a_{2}^{i}) \} 
= \partial_{i} \partial_{j} 
(\partial_{i} \partial_{j} a_{11}^{ij} 
- \partial_{i} \partial_{j} a_{11}^{ij})
= 0$. 
Hence $(\partial_{i}^{2} - \partial_{j}^{2}) \phi_{ij} = 0$. 
It follows that 
$\phi_{ij}(x_{i}, x_{j}) 
= u_{ij}^{+}{}^{''} (x_{i} + x_{j}) 
- u_{ij}^{-}{}^{''} (x_{i} - x_{j})$, 
with suitable functions $u_{ij}^{\pm}(t) = u_{ji}^{\pm}(\pm t)$. 

Now, let 
\begin{equation*}
\bar{R} 
= R 
- \sum_{1 \leq i < j \leq n} 
(u_{ij}^{+}(x_{i} + x_{j}) + u_{ij}^{-} (x_{i} - x_{j})) 
- \sum_{1 \leq i < j < k \leq n} c_{ijk} x_{i} x_{j} x_{k}. 
\end{equation*}
This function satisfies the relation 
$\partial_{i} \partial_{j} \bar{R} = 0$ for any $i < j$. 
This implies that $\bar{R} (x)$ is a sum of one variable
functions in $x_{i}$ ($1 \leq i \leq n$). 
Thus, we have proved the following lemma. 
\begin{lemma}\label{lemma:BD-2} 
There exist constants $c_{ijk}$ and one variable functions
$u_{ij}^{\pm}$ and $v_{i}$ such that 
\begin{align*}
R(x) &= 
\sum_{1 \leq i < j \leq n} 
(u_{ij}^{+}(x_{i} + x_{j}) + u_{ij}^{-} (x_{i} - x_{j})) 
+ \sum_{i=1}^{n} v_{i}(x_{i}) 
\\
& \qquad \qquad 
+ \sum_{1 \leq i < j < k \leq n} c_{ijk} x_{i} x_{j} x_{k}. 
\end{align*}
If $n = 2$, the last term is ignored. 
\end{lemma} 
Note that we may assume 
$u_{ij}^{\pm} (t) - C_{ij}^{\pm} t^{-2}$ and 
$v_{i} (t) - C_{i} t^{-2}$ to be real analytic at $t = 0$, 
as $R(x)$ is given by \eqref{eq:BD-2}.  

Let $\tilde{a}_{2}^{i}$ and $\tilde{a}_{11}^{ij}$ be functions defined
as 
\begin{align*}
a_{2}^{i} 
&= 
\tilde{a}_{2}^{i} 
- \sum_{j < k \atop \not= i} 
\{u_{jk}^{+}(x_{j} + x_{k}) + u_{jk}^{-} (x_{j} - x_{k})\}
- \sum_{j \not= i} v_{j}(x_{j}) 
\\
& \qquad \qquad 
- \sum_{j < k < l \atop \not= i} 
c_{jkl} x_{j} x_{k} x_{l}, 
\\
a_{11}^{ij} 
&= 
\tilde{a}_{11}^{ij} 
- 
u_{ij}^{+}(x_{i} + x_{j}) + u_{ij}^{-}(x_{i} - x_{j}) 
\\
& \qquad \qquad 
- \frac{1}{2} (x_{i}^{2} + x_{j}^{2}) 
\sum_{k \not= i, j} c_{ijk} x_{k} 
+ \frac{1}{3} \sum_{k \not= i, j} c_{ijk} x_{k}^{3}. 
\end{align*}
We can easily show that $\tilde{a}_{2}^{i}$ and
$\tilde{a}_{11}^{ij}$ satisfy the conditions in 
Lemma~\ref{lemma:BD-1}~(2). 

Now, note that the condition ${}^{t}P = P$ is equivalent to 
\begin{align}
a_{1}^{i} 
&= 
\partial_{i} a_{2}^{i} 
+ 
\frac{1}{2} 
\sum_{j \not= i} \partial_{j} a_{11}^{ij}
\notag \\
&= 
- \frac{1}{2} 
\sum_{j \not= i} 
\{u_{ij}^{+}{}'(x_{i} + x_{j}) 
+ u_{ij}^{-}{}'(x_{i} - x_{j}) 
- \partial_{j} \tilde{a}_{11}^{ij} \}
- \sum_{j < k \atop \not= i} c_{ijk} x_{j} x_{k}. 
\label{eq:BD-18}
\end{align}
Next, the coefficient of
$\partial_{i}$ in the relation $[L, P] = 0$ implies 
\begin{equation*} \label{eq:BD-19}
-2 \partial_{i} a_{0} 
= 
2 \sum_{j \not= i} \partial_{i} \partial_{j}^{2} R
+ 2 a_{2}^{i} \partial_{i} R 
+ \sum_{j \not= i} a_{11}^{ij} \partial_{j} R
+ \Delta a_{1}^{i}. 
\end{equation*}
Using this, we find that the compatibility condition 
$\partial_{j} (\partial_{i} a_{0}) 
= \partial_{i} (\partial_{j} a_{0})$ is
equivalent to 
\begin{align}
& 3 (\partial_{j} a_{11}^{ij} \partial_{j} R 
- \partial_{i} a_{11}^{ij} \partial_{i} R) 
+ 2 (a_{2}^{i} - a_{2}^{j}) \partial_{i} \partial_{j} R 
+ a_{11}^{ij} (\partial_{j}^{2} - \partial_{i}^{2}) R 
\label{eq:BD-20} \\
& \quad + \sum_{k \not= i,j} 
(\partial_{j} a_{11}^{ik} - \partial_{i} a_{11}^{jk}) \partial_{k} R 
+ \sum_{k \not= i,j} 
(a_{11}^{ik} \partial_{j} \partial_{k} R 
- a_{11}^{jk} \partial_{i} \partial_{k} R) 
= 0.
\notag 
\end{align}
Here, we have used \eqref{eq:BD-5} and the relations 
$\partial_{i} \partial_{j} 
(\partial_{i}^{2} - \partial_{j}^{2}) R = 0$ and 
$\Delta (\partial_{i} a_{1}^{j} - \partial_{j} a_{1}^{i}) = 0$. 
The last of these is a consequence of \eqref{eq:BD-18}. 
From \eqref{eq:BD-2}, 
it is seen that only the term 
$2 (a_{2}^{i} - a_{2}^{j}) \partial_{i} \partial_{j} R$ 
can have poles of order four at $x_{i} \pm x_{j} = 0$. 
Therefore, taking 
$\lim_{x_{j} \rightarrow \mp x_{i}} 
((x_{i} \pm x_{j})^{4} \times \eqref{eq:BD-20})$, 
we obtain 
\begin{align}
& C_{ij}^{-}
\bigg\{
\sum_{k \not= i, j} 
(u_{ik}^{+} (x_{i} + x_{k}) 
+ u_{ik}^{-} (x_{i} - x_{k})
- u_{jk}^{+} (x_{i} + x_{k}) 
- u_{jk}^{-} (x_{i} - x_{k}) 
)
\label{eq:BD-21} \\
& \qquad
+ v_{i}(x_{i}) - v_{j}(x_{i}) 
+ x_{i} \sum_{k < l \atop \not= i,j} 
(c_{ikl} - c_{jkl}) x_{k} x_{l} 
+ (\tilde{a}_{2}^{i} - \tilde{a}_{2}^{j})|_{x_{j} = x_{i}} 
\bigg\}
= 0,  
\notag 
\end{align}
and
\begin{align}
& C_{ij}^{+}
\bigg\{
\sum_{k \not= i, j} 
(u_{ik}^{+} (x_{i} + x_{k}) 
+ u_{ik}^{-} (x_{i} - x_{k})
- u_{jk}^{+} (-x_{i} + x_{k}) 
- u_{jk}^{-} (-x_{i} - x_{k}) 
)
\label{eq:BD-22} \\
& \qquad
+ v_{i}(x_{i}) - v_{j}(-x_{i}) 
+ x_{i} \sum_{k < l \atop \not= i,j} 
(c_{ikl} + c_{jkl}) x_{k} x_{l} 
+ (\tilde{a}_{2}^{i} - \tilde{a}_{2}^{j})|_{x_{j} = -x_{i}} 
\bigg\}
= 0.  
\notag 
\end{align}
Moreover, because only the terms $a_{11}^{ij} \partial_{j}^{2} R$ and 
$a_{11}^{ik} \partial_{j} \partial_{k} R$ can have poles of order
four at $x_{j} \pm x_{k} = 0$, 
taking 
$\lim_{x_{k} \rightarrow \mp x_{j}} 
((x_{j} \pm x_{k})^{4} \times \eqref{eq:BD-20})$, 
we obtain 
\begin{align}
& C_{jk}^{\pm} 
\bigg\{
u_{ij}^{+}(x_{i} + x_{j}) - u_{ij}^{-}(x_{i} - x_{j}) 
\pm u_{ik}^{+}(x_{i} \mp x_{j}) 
\mp u_{ik}^{-}(x_{i} \pm x_{j}) 
\label{eq:BD-23} \\
& \qquad + 
\sum_{l \not= i,j,k} (c_{ijl} \pm c_{ikl}) 
\left(
\frac{1}{2} (x_{i}^{2} + x_{j}^{2}) x_{l} 
- \frac{1}{3} x_{l}^{3}
\right) 
- (\tilde{a}_{11}^{ij} \pm \tilde{a}_{11}^{ik})
|_{x_{k} = \mp x_{j}}
\bigg\}
= 0. 
\notag 
\end{align}
Finally, because only the term $a_{11}^{ij} \partial_{i}^{2} R$ can
have a pole of order four at $x_{i} = 0$, 
taking 
$\lim_{x_{i} \rightarrow 0} 
(x_{i}^{4} \times \eqref{eq:BD-20})$, 
we obtain 
\begin{align}
& C_{i}
\bigg\{
u_{ij}^{+}(x_{j}) - u_{ij}^{-}(- x_{j}) 
+ 
\sum_{k \not= i,j} c_{ijk} 
\left(
\frac{1}{2} x_{j}^{2} x_{k} 
- \frac{1}{3} x_{k}^{3} 
\right) - \tilde{a}_{11}^{ij} |_{x_{i} = 0} 
\bigg\}
= 0. 
\label{eq:BD-25} 
\end{align}

Next, the limits 
$\lim_{x_{k} \rightarrow \mp x_{i}} 
((x_{i} \pm x_{k})^{2} \times \eqref{eq:BD-21})$, 
$\lim_{x_{k} \rightarrow \mp x_{i}} 
((x_{i} \pm x_{k})^{2} \times \eqref{eq:BD-22})$, 
$\lim_{x_{i} \rightarrow 0} 
(x_{i}^{2} \times \eqref{eq:BD-21} \mbox{ or } 
\eqref{eq:BD-22})$ and 
$\lim_{x_{j} \rightarrow 0} 
(x_{j}^{2} \times \eqref{eq:BD-25})$ 
give 
\begin{align}
& C_{ij}^{-} (C_{ik}^{\pm} - C_{jk}^{\pm}) = 0, 
\label{eq:BD-26}
\\ 
& C_{ij}^{+} (C_{ik}^{\pm} - C_{jk}^{\mp}) = 0, 
\label{eq:BD-27} 
\\
& C_{ij}^{\pm} (C_{i} - C_{j}) = 0, 
\label{eq:BD-28} 
\\
& C_{i} (C_{ij}^{+} - C_{ij}^{-}) = 0  
\label{eq:BD-29} 
\end{align}
for $i \not= j \not= k \not= i$, 
because $u_{ij}^{\pm}(t)$ and $v_{i}(t)$ can have poles of order two
at $t = 0$. 

Because $\mathcal{H}$ is not empty, 
at least one of $C_{ij}^{\pm}$ ($1 \leq i < j \leq n$) or 
$C_{i}$ ($1 \leq i \leq n$) is not zero. 
If all the $C_{ij}^{\pm}$ are zero, then $\mathcal{H}$ is
divided into nonempty orthogonal subsets. 
However, this contradicts the condition (I2). 
Therefore, applying an appropriate coordinate transformation, 
we are able to realize the condition $C_{12}^{-} \not= 0$. 
Then, \eqref{eq:BD-26} and \eqref{eq:BD-28} imply 
$C_{1i}^{\pm} = C_{2i}^{\pm}$ for $i \geq 3$ and 
$C_{1} = C_{2}$. 

If $n = 2$ and $C_{1} = C_{2} = 0$, then 
$\mathcal{H} = \{e_{1} + e_{2}, e_{1} - e_{2}\}$, 
but this contradicts the condition (I2). 
Therefore, $C_{1} = C_{2} \not= 0$. 
Then, from \eqref{eq:BD-29}, 
we obtain $C_{12}^{+} = C_{12}^{-} \not= 0$. 
Therefore, $\mathcal{H}$ coincides with the positive system
of the root system of type $B_{2}$. 

Now, assume $n \geq 3$. 
Then, using the same argument as in the $A_{n-1}$-case, 
we can show that $C_{ij}^{+}$, $C_{ij}^{-}$ and $C_{i}$ are all 
independent of $i$ and $j$. 
We write $C^{\pm} := C_{ij}^{\pm}$ and $C := C_{i}$. 
If $C^{+} \not= C^{-}$, then $C = 0$ and $C^{+} = 0$, as found from 
\eqref{eq:BD-27} and \eqref{eq:BD-29}. 
This implies that the hyperplane arrangement $\mathcal{H}$
is an $A_{n-1}$-type positive system, which contradicts our
assumption $W = W(B_{n})$ or $W(D_{n})$. 
Therefore, $C^{+} = C^{-} \not= 0$. 
If $C = 0$, then $\mathcal{H}$ is of type $D_{n}$, 
and if $C \not= 0$ it is of type $B_{n}$. 

Combining the above results, 
we have proved the following proposition. 
\begin{proposition} 
Under the assumptions made in this section, 
the hyperplane arrangement $\mathcal{H}$ coincides with
the positive root system of type $B_{n}$ or type $D_{n}$. 
Moreover, the parameters $C_{\alpha}$ in
\eqref{eq:Schrodinger} are $W$-invariant. 
\end{proposition}

\begin{lemma}\label{lemma:BD-4} 
For any $1 \leq i < j < k \leq n$, we have 
\[
c_{ijk} = 0. 
\]
\end{lemma} 
\begin{proof} 
We can assume $n \geq 3$. 
If $\mathcal{H}$ is of type $B_{n}$, 
then the fact that $C_{i} \not= 0$, 
the relation obtained by applying $\partial_{k}^{3}$ to
\eqref{eq:BD-25}, 
and Lemma~\ref{lemma:BD-1} together imply that $c_{ijk} = 0$. 
If $\mathcal{H}$ is of type $D_{n}$ ($n \geq 4$), 
then the fact that $C_{jk}^{\pm} \not= 0$, the relation obtained by
applying $\partial_{l}^{3}$ to \eqref{eq:BD-23}, 
and Lemma~\ref{lemma:BD-1} together imply that 
$c_{ijl} \pm c_{ikl} = 0$, and hence $c_{ijk} = 0$ for any $i, j, k$. 
\end{proof} 

\begin{lemma} 
$\tilde{a}_{2}^{i}$ and $\tilde{a}_{11}^{ij}$ can be expressed
as follows: 
\begin{align}
\tilde{a}_{2}^{i} 
&= 
- \frac{1}{2} \sum_{j \not= i} \alpha_{ij} x_{j}^{2} 
- \sum_{j \not= i} \beta_{ij}(i) x_{j} + \delta_{i}, 
\label{eq:BD-30} 
\\
\tilde{a}_{11}^{ij} 
&= \alpha_{ij} x_{i} x_{j} + \beta_{ij}(i) x_{i} 
+ \beta_{ij}(j) x_{j} + \gamma_{ij}. 
\label{eq:BD-31}
\end{align}
Here, $\alpha_{ij}, \dots, \delta_{i}$ can be any constants
satisfying $\sum_{i=1}^{n} \delta_{i} = 0$. 
\end{lemma} 
\begin{proof} 
First, we show that $\tilde{a}_{11}^{ij}$ can be
expressed as \eqref{eq:BD-31}. 

If $\mathcal{H}$ is of type $B_{2}$, 
this is clear from Lemma~\ref{lemma:BD-1}. 

If $\mathcal{H}$ is of type $B_{3}$, 
we may assume that $\tilde{a}_{11}^{ij}$ is given by 
\[
\tilde{a}_{11}^{ij} 
= \alpha_{ij} x_{i} x_{j} + \varphi_{ij}^{i} (x_{k}) x_{i} 
+ \varphi_{ij}^{j} (x_{k}) x_{j} + \varphi_{ij}(x_{k}), 
\]
with suitable polynomials $\varphi_{ij}^{i}$, $\varphi_{ij}^{j}$ and
$\varphi_{ij}$, by Lemma~\ref{lemma:BD-1}. 
Then, \eqref{eq:BD-25} and Lemma~\ref{lemma:BD-4} imply 
\begin{equation*}
\partial_{k} \tilde{a}_{11}^{ij}|_{x_{i} = 0} = 0 
\quad 
\mbox{for $k \not= i,j$},  
\end{equation*}
so that 
$\partial_{k} \varphi_{ij}^{j} = \partial_{k} \varphi_{ij} = 0$. 
Therefore, $\tilde{a}_{11}^{ij}$ can be expressed as in
\eqref{eq:BD-31}. 

If $\mathcal{H}$ is of type $B_{n}$ or $D_{n}$ ($n \geq 4$), 
then we find 
\[
\partial_{l} (\tilde{a}_{11}^{ik} \pm \tilde{a}_{11}^{ij})
 |_{x_{k} = \mp x_{j}} 
= 0
\]
from \eqref{eq:BD-23} and Lemma~\ref{lemma:BD-4}. 
Then, using the same argument as in the $B_{3}$ case, 
we can show that
$\tilde{a}_{11}^{ij}$ is expressed as in \eqref{eq:BD-31}. 

Next, note that because 
\[
\partial_{j} \tilde{a}_{2}^{i} 
= - \partial_{i} \tilde{a}_{11}^{ij} 
= - \alpha_{ij} x_{j} - \beta_{ij}(i), 
\]
$\tilde{a}_{2}^{i}$ can be expressed as in \eqref{eq:BD-30}. 
By subtracting appropriate constant multiple of $L$ from
$P$,
we can realize the condition $\sum_{i=1}^{n} \delta_{i} = 0$. 
\end{proof} 

Now, let 
\begin{align*}
\tilde{u}_{ij}^{\pm} (t) 
&= u_{ij}^{\pm} (t) 
- \frac{\alpha_{ij}}{4} t^{2} 
- 
\frac{\beta_{ij}(j) \pm \beta_{ij}(i)}{2} t 
\mp \frac{\gamma_{ij}}{2} 
\quad 
\mbox{and}  
\\
\tilde{v}_{i} (t) 
&= 
v_{i} (t) 
+ \frac{1}{2} 
\left(\sum_{j \not= i} \alpha_{ij} \right)
t^{2} 
+ 
\left(\sum_{j \not= i} \beta_{ij}(j) \right)
t 
+ \delta_{i}. 
\end{align*}
Then, we have 
\begin{align}
R(x) &= 
\sum_{i < j} 
\{ 
\tilde{u}_{ij}^{+}(x_{i} + x_{j}) 
+ 
\tilde{u}_{ij}^{-}(x_{i} - x_{j}) 
\} 
+ \sum_{i} 
\tilde{v}_{i}(x_{i}), 
\notag \\
a_{2}^{i} 
&= 
- \sum_{j < k \atop \not= i} 
\{
\tilde{u}_{jk}^{+}(x_{j} + x_{k}) 
+ 
\tilde{u}_{jk}^{-}(x_{j} - x_{k}) 
\} 
- \sum_{j \not= i} 
\tilde{v}_{j}(x_{j}), 
\label{eq:BD-101}
\\
a_{11}^{ij} 
&= 
- \tilde{u}_{ij}^{+} (x_{i} + x_{j}) 
+ \tilde{u}_{ij}^{-} (x_{i} - x_{j}). 
\label{eq:BD-102}
\end{align}
Hence, we can realize the relation 
$\tilde{a}_{2}^{i} = \tilde{a}_{11}^{ij} = 0$ by appropriately
choosing $u$ and $v$. 

\begin{theorem}\label{theorem:BD-final}  
There exist even functions $u(t)$ and $v(t)$ such that 
\begin{equation}\label{eq:BD-33}
R(x) 
= 
\sum_{1 \leq i < j \leq n} 
\{u(x_{i} + x_{j}) + u(x_{i} - x_{j}) \} 
+ \sum_{i = 1}^{n} v(x_{i}). 
\end{equation} 
\end{theorem} 
\begin{proof} 
First, we prove the assertion for the $B_{2}$ case. 
From \eqref{eq:BD-21} and \eqref{eq:BD-22}, we have 
\[v_{2}(t) = v_{1}(t) = v_{2}(-t), 
\]
which implies that $v_{1} = v_{2}$ and that they are even
functions. 
Then, from \eqref{eq:BD-25}, we obtain 
\[
u_{12}^{+}(t) = u_{12}^{-}(-t) 
\qquad \mbox{and} \qquad 
u_{21}^{+}(t) = u_{21}^{-}(-t). 
\]
Therefore, because $u_{21}^{\pm}(t) = u_{12}^{\pm}(\pm t)$, 
we find that $u_{12}^{+}(t) = u_{12}^{-}(t)$, and they are even
functions. 

Now, consider the $B_{n}$ and $D_{n}$ cases for $n \geq 3$. 
In this case, from \eqref{eq:BD-23}, 
we have 
\begin{equation*}
u_{ij}^{+}(x_{i} + x_{j}) - u_{ij}^{-}(x_{i} - x_{j}) 
\pm u_{ik}^{+}(x_{i} \mp x_{j}) 
\mp u_{ik}^{-}(x_{i} \pm x_{j}) 
= 0, 
\end{equation*}
which imply 
\begin{align*}
& u_{ij}^{+}(s) - u_{ik}^{-}(s) 
= 
u_{ij}^{-}(t) - u_{ik}^{+}(t), 
\\
& u_{ik}^{+}(s) - u_{ik}^{-}(s) 
= 
-(u_{ik}^{+}(t) - u_{ik}^{-}(t)). 
\end{align*}
Hence, 
$u_{ij}^{+}(t) - u_{ik}^{-}(t)$ is a constant, 
and $u_{ik}^{+}(t) = u_{ik}^{-}(t)$. 
Therefore, there exist constants 
$p_{ij}$ and a function $u(t)$ such that 
\begin{equation}\label{eq:BD-35}
u_{ij}^{+}(t) = u_{ij}^{-}(t) = u(t) + p_{ij}. 
\end{equation}
Note that $u(t)$ is an even function, because 
\[
u(t) - u(-t) 
= 
u_{ij}^{+}(t) - u_{ji}^{-}(-t) 
= 
u_{ij}^{+}(t) - u_{ij}^{-}(t) = 0. 
\]
Here, we have used $u_{ij}^{\pm}(t) = u_{ji}^{\pm}(\pm t)$. 
Moreover, because $u(t)$ is fixed up to an arbitrary constant, 
we can realize the condition 
\begin{equation}\label{eq:BD-36} 
\sum_{i<j} p_{ij} = 0.
\end{equation} 
Next, substituting \eqref{eq:BD-35} into \eqref{eq:BD-21}, 
we obtain 
$2 \sum_{k \not= i, j} (p_{ik} - p_{jk}) 
+ v_{i}(x_{i}) - v_{j}(x_{i}) = 0$, 
or 
\[
v_{i}(t) + 2 \sum_{k \not= i} p_{ik} 
= 
v_{j}(t) + 2 \sum_{k \not= j} p_{jk}. 
\]
This implies that the function $v_{i}(t) + 2 \sum_{k \not= i} p_{ik}$
is independent of $i$, 
and we write it as $v(t)$. 

From \eqref{eq:BD-36}, 
we have 
$\sum_{i} v_{i}(x_{i}) 
- 
\sum_{i} v(x_{i}) 
= - 2 \sum_{i} \sum_{k \not= i} p_{ik} 
= 0$ 
and 
$\sum_{i<j} 
\{ 
u_{ij}^{+}(x_{i} + x_{j}) + u_{ij}^{-}(x_{i} - x_{j}) 
\}
= 
\sum_{i<j} 
\{ 
u(x_{i} + x_{j}) + u(x_{i} - x_{j}) 
\}$. 
Hence $R(x)$ can be expressed as in \eqref{eq:BD-33}. 

Finally, because $u(t)$ is an even function, 
\eqref{eq:BD-22} implies that $v(t)$ is also an even function. 
\end{proof} 

\begin{remark}\label{rem:BD-type last}
Applying \eqref{eq:BD-35} and the relation 
$v_{i}(t) = v(t) - 2 \sum_{k \not= i} p_{ik}$ 
to \eqref{eq:BD-101} and \eqref{eq:BD-102}, 
we obtain 
\begin{align*}
a_{2}^{i} 
&= 
- \sum_{j < k \atop \not= i} 
\{u(x_{j} + x_{k}) + u(x_{j} - x_{k})\} 
- \sum_{j \not= i} v(x_{j}), 
\\
a_{11}^{ij} 
&= 
- u(x_{i} + x_{j}) + u(x_{i} - x_{j}). 
\end{align*}
Therefore, the functional equation \eqref{eq:BD-20} is identical to 
that studied by Ochiai, Oshima and Sekiguchi in \cite{OO, OS}. 
This equation has been completely solved, 
and the solutions are given in Theorem~1 of \cite{OOS}. 
\end{remark} 
%




\begin{thebibliography}{99} 

\bibitem{BC} Burchnall, J.L. and Chaundy, T.W.: 
Commutative ordinary differential operators. 
Proc. London Math. Soc. (2) {\bf 21} (1922), 420--440. 

\bibitem{Ca} Calogero, F.: 
Solution of the one-dimensional $n$-body problem with quadratic
  and/or inversely quadratic pair potentials. 
J. Math. Phys. {\bf 12} (1971), 419--436. 

\bibitem{Ch} Chalykh, O.A.: 
Additional integrals of the generalized quantum Calogero-Moser
  problem. 
Theoret. and Math. Phys. {\bf 109} (1996), 1269--1273. 

\bibitem{CFV} Chalykh, O., Feigin, M. and Veselov, A.: 
New integrable generalization of Calogero-Moser quantum problem. 
J. Math. Phys. {\bf 39} (1998), 695--703. 

\bibitem{CV} Chalykh, O.A. and Veselov, A.P.: 
Commutative rings of partial differential operators and Lie
  algebras. 
Comm. Math. Phys. {\bf 126} (1990), 597--611. 

\bibitem{M} Moser, J.: 
Three integrable Hamiltonian systems connected with isospectral
  deformations. 
Advances in Math. {\bf 64} (1975), 197--220. 

\bibitem{OO} Ochiai, H. and Oshima, T.: 
Commuting differential operators of type $B_2$. 
S\=urikaisekikenky\=usho K\=oky\=uroku, {\bf 1171} (2000), 36--67.  

\bibitem{OOS} Ochiai, H., Oshima, T. and Sekiguchi, H.: 
Commuting families of symmetric differential operators. 
Proc. Japan Acad., {\bf 70 A} (1994), 62--66. 

\bibitem{OP2} Olshanetsky, M.A. and Perelomov, A.M.: 
Quantum integrable systems related to Lie algebras. 
Phys. Rep. {\bf 94} (1983), 313--404. 

\bibitem{OS} Oshima, T. and Sekiguchi, H.: 
Commuting families of differential operators invariant under 
the action of a Weyl group. 
J. Math. Sci. Univ. Tokyo, {\bf 2} (1995), 1--75. 

\bibitem{T} Taniguchi, K.: 
On uniqueness of commutative rings of Weyl group invariant
  differential operators. 
Publ. RIMS, Kyoto Univ. \textbf{33} (1997), 257--276.

\bibitem{VFC} Veselov, A.P., Feigin, M. and Chalykh,
O.A.: 
New integrable deformations of the quantum Calogero-Moser
problem. 
Russian Math. Surveys {\bf 51} (1996), No. 3, 573-574. 

\bibitem{VSC} Veselov, A.P., Styrkas, K.L. and Chalykh, O.A.: 
Algebraic integrability for the Schr\"{o}dinger equation, 
and groups generated by reflections. 
Theor. Math. Phys. {\bf 94} (1993), 182--197. 



\end{thebibliography}
\end{document}